\documentclass[twocolumn,english]{revtex4}
\usepackage[latin1]{inputenc}
\usepackage{graphicx}
\usepackage{amssymb}

\begin{document}

\title{Testing non-classical theories of electromagnetism with ion interferometry}

\author{B. Neyenhuis, D. Christensen, and D. S. Durfee}

\affiliation{Brigham Young University, Department of Physics and Astronomy, Provo,
UT 84602}


\date{19 September 2007}

\begin{abstract}
We discuss using a table-top ion interferometer to search for deviations
from Coulomb's inverse-square law. Such deviations would result from
non-classical effects such at a non-zero photon rest mass. We discuss
the theory behind the proposed measurement, explain which fundamental,
experimentally controllable parameters are the relevant figures of
merit, and calculate the expected performance of such a device in
terms of these parameters. The sensitivity to deviations in the exponent
of the inverse-square law is predicted to be a few times $10^{-22}$,
an improvement by five orders of magnitude over current experiments.
It could measure a non-zero photon rest mass smaller than $9\times10^{-50}$
grams, nearly $100$ times smaller than current laboratory experiments. 
\end{abstract}

\maketitle

The experimental search for deviations from current theories will
eventually lead to the next, more fundamental theory of physics. Such
studies challenge the Standard Model and give insight into the form
of the underlying, more elemental theory. Coulomb's inverse-square
law is the foundational law in electrostatics. Gauss's Law and Maxwell's
equations are built upon this law and the principle of superposition.
Precision tests of this law are essential to push forward our understanding
of electromagnetism and its relation to the other forces.

Detection of any deviation from Coulomb's law would have far-reaching
implications. Maxwell's equations and much of the Standard Model would
have to be modified. The notion that absolute electrostatic potential
is arbitrary would have to be abandoned, along with many other tenets
of classical electromagnetism. Inverse-square-law violation would
suggest a finite range for the electromagnetic force, implying a non-zero
photon rest mass \cite{Goldhaber71,Tu2005,Jackson1975}. Consequences
of massive photons include a frequency-dependent velocity of light
in vacuum and electromagnetic waves with a longitudinal component
of polarization \cite{Greiner1996}. Several grand-unification theories
include massive photons \cite{Arkani1998,Kostelecky1991}, and further
tests of the inverse-square law can help confirm or disprove them.

Several studies have searched for consequences of massive photons
rather than testing the inverse-square law directly \cite{Tu2005}.
These studies involve many assumptions about the nature of interstellar
space and the sources of the measured light waves. It is therefore
necessary to verify these results with laboratory experiments where
variables can be better controlled %
\footnote{The measurements in \cite{Luo2003,Lakes1998} are sometimes erroneously
considered laboratory experiments. But to obtain the photon rest mass
from this type of measurement the cosmic vector potential must be
guessed, making the experiment more like model-dependent astronomical
studies. %
}. And while the possibility of a massive photon supplies additional
motivation and provides a common parameter to compare experiments,
it is possible that Coulomb's law is violated for reasons unrelated
to photon rest mass. Only an experiment which specifically measures
the inverse-square law would be sensitive to these effects.

Although Coulomb's law has been tested many times over the last two
and a half centuries \cite{Robinson1769,Cavendish1773,Maxwell,Jackson1975,Williams1971,Crandall83},
this subject has seen little progress in the last three decades. The
smallest laboratory-based limit on the photon rest mass was reported
24 years ago \cite{Crandall83}. In this experiment an alternating
voltage was applied between two conducting shells, and the induced
voltage between the outer of the two and a third shell was measured
with solid-state electronics. This measurement improved upon the best
previous measurement, 12 years old at the time \cite{Williams1971},
by only a factor of 2.5. In this paper we show that it should be possible
to revitalize this key field of study and improve sensitivity by orders
of magnitude using a new approach --- charged particle matter-wave
interferometry. 

\begin{figure}
\begin{centering}
\includegraphics[width=8.4cm,keepaspectratio]{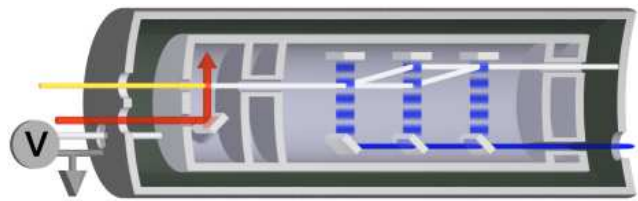}
\end{centering}

\caption{(Color online) A cut-away cartoon of the proposed experiment. The
diagram is not to scale, and some dimensions have been greatly exaggerated
for visibility.\label{cap:Schematic} }
\end{figure}

In the proposed experiment, a possible Coulomb's-law violating electric
field inside of a conducting shell is measured with an ion interferometer.
As shown in Fig.\ \ref{cap:Schematic}, ions travel through a conducting
cylinder nested inside of a second cylinder. The outer conductor is
grounded, and a time-varying voltage is applied to the inner conductor.
A slow beam of atoms passes through small holes in the conductors.
The atoms are ionized with a laser beam, shown as an arrow in the
figure, and pass through three gratings to form a Mach-Zehnder interferometer.
If an electric field is present in the inner conductor, the two interferometer
arms will pass through different potentials, resulting in a relative
phase shift. Using optical gratings would allow state-selective readout
and avoid drawbacks of physical gratings \cite{MetalizedGratings},
including charge build-up and image charges in the gratings. Using
Raman transitions \cite{Gustavson97} would allow precise control
of grating phases. 

To calculate the sensitivity that could be achieved, we start with
a modified version of Laplace's equation derived from the Proca action
for massive photons:

\begin{equation}
\bigtriangledown^{2}\phi-\mu_{\gamma}^{2}\phi=0.\label{eq:LaplaceP}\end{equation}
In this equation $\phi$ is the scalar electrostatic potential, and
$\mu_{\gamma}=m_{\gamma}c/\hbar$, where $m_{\gamma}$ is the photon
rest mass, $\hbar$ is Planck's constant divided by $2\pi$, and $c$
is the canonical speed of light in vacuum.

In the limit as $\mu_{\gamma}\rightarrow0$, Eq.\ \ref{eq:LaplaceP}
becomes Laplace's equation. For a spherically symmetric system, Laplace's
equation has the familiar solutions $\phi(r)=A/r$ and $\phi(r)=B$,
where $A$ and $B$ are constants. The $A/r$ solution is the classical
point-charge potential. The constant $B$ solution allows us to arbitrarily
define a point to be at zero potential without changing the fields
described by the potential. If $\mu_{\gamma}\neq0$, the solutions
for a spherically symmetric system are a Yukawa potential $\phi(r)=(A/r)\exp(-\mu_{\gamma}r)$,
and an exponentially growing solution $\phi(r)=(B/r)\exp(\mu_{\gamma}r)$.
The Yukawa-potential solution lets us interpret $1/\mu_{\gamma}$
as an effective range of the Coulomb force. Without a constant solution,
absolute potential has physical significance and we are no longer
free to arbitrarily choose where $\phi$ equals zero.

\begin{figure}
\begin{centering}
\includegraphics[width=8.5cm,height=4cm,keepaspectratio]{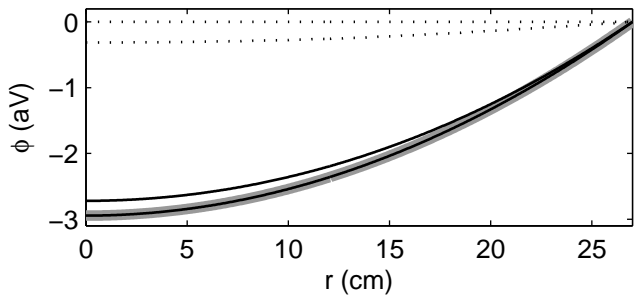}
\end{centering}

\caption{Calculations of potentials in a 2.6 m long, 27 cm radius tube held
at 200 kV. The calculation assumes $m_{\gamma}$$=1\times10^{-50}$
grams. Potentials are plotted vs.\ the radial distance from the tube
axis. The black lines are the deviation from the classical potential
at axial distances of zero (lower line) and one meter (upper line)
from the middle of the tube, plotted on top of a thick gray line representing
the deviation for an infinite tube. The dotted lines show the calculated
classical fringing-field potentials at the same locations multiplied
by $10^{35}$ to make them visible on this scale. \label{cap:NumericalStudy} }
\end{figure}

Due to the elongated geometry of the proposed experiment, we will
approximate the finite inner conductor with an infinitely long tube.
Numerical and analytical studies have verified that this is a good
approximation for reasonably long tubes (see Fig.\ \ref{cap:NumericalStudy}
and \cite{Christensen06nc}). For a system with no angular or longitudinal
dependence, solutions to Eq.\ \ref{eq:LaplaceP} are the zeroth-order
modified Bessel functions. Applying the conditions that $\phi(r)$
must equal the applied voltage when $r=R$ (the radius of the tube),
we determine that to lowest order in $\mu_{\gamma}$ the potential
inside the inner tube is given by

\begin{equation}
\phi(r)\approx(V+V_{g})\left[1+\frac{\mu_{\gamma}^{2}}{4}(r^{2}-R^{2})\right].\end{equation}
Here $V$ is the voltage applied to the inner tube relative to the
outer tube, and $V_{g}$ is the unknown voltage of the outer, grounded
tube.

Rather than absolute potential, the interferometer will measure the
potential difference between the two arms. Each of the arms in Fig.\
\ref{cap:Schematic} consists of one horizontal and one diagonal segment.
The diagonal segments can be neglected because they both pass through
identical potentials inducing equal phase shifts. The horizontal segments,
however, travel through different potentials. Assuming that the two
horizontal segments are a distance $r_{0}$ and $r_{0}+s$ from the
center of the tube, the potential difference between them is \begin{equation}
\Delta\phi=\phi(r_{0}+s)-\phi(r_{0})\approx\frac{\mu_{\gamma}^{2}}{4}\left(V+V_{g}\right)\left(s^{2}+2r_{0}s\right).\end{equation}

If $\tau$ is the time that it takes the ions to travel the length
of the horizontal segments, and $e$ is the ion charge, the interferometer
phase $\Phi$ is given by \begin{equation}
\Phi\approx\frac{e\mu_{\gamma}^{2}}{4}\left(V+V_{g}\right)\left(s^{2}+2r_{0}s\right)\frac{\tau}{\hbar}+\Phi_{0},\end{equation}
where $\Phi_{0}$ is the phase when $V+V_{g}=0$. This term includes
all phase shifts which are not dependent on the absolute potential,
such as those due to patch charges, imbalanced interferometer arms,
etc. Although $V_{g}$ and $\phi_{0}$ are unknown, one could change
the potential $V$ by an amount $\Delta V$ and look for a correlated
change in the interferometer phase. Because the $\sim700\,\mu\mbox{F}$
capacitance of the Earth is very large compared to the $\sim1.6\mbox{ nF}$
capacitance of the proposed conductors, $V_{g}$ will remain roughly
constant as $V$ is changed, and the difference in phase due to the
potential change will be

\begin{equation}
\Delta\Phi\approx\frac{e\Delta V\mu_{\gamma}^{2}\tau}{4\hbar}\left(s^{2}+2r_{0}s\right).\label{eq:PhaseShift}\end{equation}
 Solving Eq.\ \ref{eq:PhaseShift} for $\mu_{\gamma}$ we can determine
the rest mass of the photon from the measured interferometer phase
shift:\begin{equation}
m_{\gamma}\approx\frac{\hbar}{c}\left[\frac{4\hbar\Delta\Phi}{e\Delta V\left(s^{2}+2r_{0}s\right)\tau}\right]^{1/2}.\label{eq:Mgamma}\end{equation}

To estimate the smallest detectable $m_{\gamma}$, it is useful to
rewrite Eq.\ \ref{eq:Mgamma} in terms of experimentally accessible
parameters. One important parameter is the velocity of the ions $v$.
A smaller velocity results in larger diffraction angles but more deflection
by stray electric fields. Fields will be extremely small inside the
tube at the locations of the gratings (see Fig.\ \ref{cap:NumericalStudy}),
but they could be much larger in the region where the ions are generated.
If we write the velocity as $v=(2eV_{s}/m)^{1/2}$ where $m$ is the
mass of the ions and $V_{s}$ is the voltage which would just bring
the ions to a stop, we can set $V_{s}$ to be several times the level
of the expected stray fields to be sure that the trajectory of the
ions is not greatly perturbed by them.

Two other important parameters are the maximum excursion of the ions
from the center of the tube, $a=r_{0}+s$, and the distance between
gratings, $L$. A larger tube radius accommodates a larger separation
$s$ and offset $r_{0}$. A larger grating separation $L$ means that
the ions will interact with the field longer ($\tau=L/v$) and results
in a greater separation of the two arms of the interferometer ($s\approx Lh/mvd$,
where $h$ is Planck's constant and $d$ is the grating period). With
these parameters in mind, we can rewrite Eq.\ \ref{eq:Mgamma} as\begin{equation}
m_{\gamma}\approx\frac{\hbar}{cL}\left[\frac{2\Delta\Phi dV_{s}}{\pi\Delta Va\left(1-Q\right)}\right]^{1/2}\label{eq:MgammaB}\end{equation}
 where the parameter $Q=s/2a=\pi L\hbar/(2emV_{s}a^{2}d^{2})^{1/2}$
shows very weak dependence on ion mass and charge --- although higher
charge and lower mass results in more precision for a given ion velocity,
this is offset by the greater velocity needed to overcome deflections
by stray fields. For arbitrary experimental parameters, $0<Q\leq1/2$.
For the parameters selected below, $Q$ is small for all possible
ion masses, ranging from $1\times10^{-2}$ for $^{1}\textrm{H}^{+}$
to $1\times10^{-3}$ for $^{133}\textrm{Cs}^{+}$, and the precision
of the experiment will not change much with the mass or charge of
the ion. 

There are practical limits on $L$ and $a$ for a table-top apparatus.
We chose $L$ to be one meter, and limited $a$ to be a conservative
25 cm. For our numerical calculations (Fig.\ \ref{cap:NumericalStudy})
we assumed a total length for the inner tube plus end-caps of 3 m,
and a tube radius of 27 cm. This gives sufficient space to limit fringing
fields, to be sure that the infinite-tube calculation is a good approximation
in the region of the interferometer, and to keep the ions in the interferometer
away from the tube surface. Only $a$, and not the outer radius of
the tube, affect the precision predicted in Eq.\ \ref{eq:MgammaB}.
So a tube with a larger radius could be used to further limit ion-surface
interactions without changing the predicted precision.

We assumed a grating period of 200 nm, about half the wavelength of
a readily-available uv diode laser. We selected a value of 400 kV
for $\Delta V$ because $\pm200\mbox{ kV}$ is within the range of
what is possible with off-the-shelf power supplies and vacuum feed-throughs.
Based on work done with atom interferometers \cite{Gustavson00},
it should be possible to detect phase-shifts as small as $10^{-4}$
radians. We set the final parameter, $V_{s}$, to 0.5 mV assuming
that voltages due to stray fields can be controlled well below this
level.

Given the assumed parameters the separation $s$ would range from
6.4 mm (for $^{1}\textrm{H}^{+}$) to 0.56 mm (for $^{133}\textrm{Cs}^{+}$),
and the ion beam would enter the apparatus at a radius $r_{0}$ ranging
from 24.4 cm ($^{1}\textrm{H}^{+}$) to 24.9 cm ($^{133}\textrm{Cs}^{+}$).
For electrons these parameters yield an $s$ larger than the radius
of the tube, making electrons a poor choice for these experimental
parameters. For a horizontal apparatus in gravity, assuming that the
ions undergo a parabolic trajectory with the peak at the location
of the center grating, the ions will fall a vertical distance ranging
from 51 $\mu$m ($^{1}\textrm{H}^{+}$) to 6.8 mm ($^{133}\textrm{Cs}^{+}$),
giving them a vertical velocity of only $1.0\times10^{-4}$ ($^{1}\textrm{H}^{+}$)
to $1.4\times10^{-2}$ ($^{133}\textrm{Cs}^{+}$) times their longitudinal
velocity. The phase shift due to gravity will be constant as the applied
voltage is changed, and will not affect the measurement.

The velocity of the ions, determined by $V_{s}$ and the mass of the
ion, ranges from 311 m/s ($^{1}\textrm{H}^{+}$) to 27 m/s ($^{133}\textrm{Cs}^{+}$).
Ions at these velocities could be generated by ionizing a slow neutral-atom
beam; a velocity of 27 m/s is a reasonable velocity for a beam of
atoms from an LVIS source \cite{LVIS}. Higher velocities are easily
obtained by accelerating the ions with a small potential. As such,
any atom that can be laser cooled could be used, all resulting in
similar sensitivity. However, lighter ions have the advantage of faster
transit, which would make it possible to modulate the voltage applied
to the tube at a higher frequency, reducing the effective bandwidth
of systematic drifts.

With these parameters we predict a sensitivity to photon rest mass
of $9\times10^{-50}$ grams, nearly two orders of magnitude smaller
than the limit in \cite{Crandall83}. In addition to photon-rest-mass
limits, following the tradition of Cavendish \cite{Cavendish1773}
it is also common to assume that the point-charge potential falls
off as $r^{-(1+\delta)}$ and to quantify inverse-square-law violations
with the small parameter $\delta$. Because the Proca treatment isn't
necessarily correct, this additional figure of merit is valuable.
Unfortunately, the $r^{-(1+\delta)}$ potential does not come from
an underlying theory. If such a theory existed, $r^{-(1+\delta)}$
would enter naturally as a solution to a modified version of Laplace's
equation. At least one more solution, one which is finite at $r=0$,
must exist. Knowing just one of the solutions is not sufficient to
determine charge distributions.

It appears that previous experiments calculated limits on $\delta$
by integrating the point-charge potential over the classical charge
distribution. But the unmodelled deviation of the true charge distribution
from the classical distribution could greatly affect the magnitude
of $\delta$. Furthermore, if the unknown equation is nonlinear, the
potential cannot be related to an integral of point charges. It is
also disturbing that in this formalism the units of the permittivity
$\epsilon_{0}$ depend on $\delta$. Ignoring these concerns, we integrated
the point-charge potential over the classical charge distribution
and predict a limit on $\delta$ of a few times $10^{-22}$ in the
proposed apparatus, an improvement of five orders of magnitude over
the value reported in \cite{Crandall83}.

In addition to higher sensitivity, the proposed device overcomes a
potential pitfall present in the most recent experiments. In these
studies \cite{Williams1971,Crandall83} a voltage between two conducting
shells was measured electronically. Any non-zero electric field would
tend to draw a charge through the electronics to cancel the field.
If $0.001$ times the charge of an electron passed through the probe
electronics, it would cancel the field due to a photon mass larger
than the reported precision. But in our scheme the only influence
the ions have on the system under test is the well-understood induction
of an image charge in the conductor. 

The largest errors in the proposed measurement are expected to be
due to inertial-force shifts \cite{Gustavson00} and ion-ion interactions.
The ion-ion interactions can be reduced by limiting the number of
ions inside the conductor at any given time --- in the limit of a
single ion at a time, this effect disappears while still affording
a count rate of tens to hundreds of ions per second. This drift can
also be reduced by using non-classical, anti-bunched ion beams. 

We performed numerical calculations and piecewise analytical solutions
to verify that fringing fields from holes in the conductor could be
made negligibly small \cite{Christensen06nc}. The calculations show
that fringing fields will be tens of orders of magnitude below the
detection limit (see Fig.\ \ref{cap:NumericalStudy}), and should
not be an issue. In these calculations the inner conductor was capped
with 20 cm-long end caps to reduce fringing fields. Because the calculations
assumed axial symmetry, the holes in the end cap were replaced with
ring-shaped apertures. As such, the calculations greatly overestimate
the size of the fringing fields. 

Drifts in patch charges \cite{Deslauriers06sa} and similar effects
should not be correlated with changes in the applied voltage, especially
considering the extremely small level of the fringing fields. The
susceptibility to stray electric fields should be no greater than
in other recent laboratory tests of Coulomb's law, and in \cite{Crandall83}
it was implied that these effects were not a limitation. Magnetic
shielding will be necessary, and magnetic fields created by the charging
and discharging of the conductors will have to be taken into account.
The effect of eddy currents could be reduced exponentially by increasing
the time between voltage reversals. And although static fields will
not affect the phase difference, large static fields will reduce fringe
contrast. Nevertheless, these difficulties are surmountable in a reasonable
experiment.

In conclusion, we have discussed the prospect of using ion interferometry
to search for violations of Coulomb's law. Calculations using reasonable
parameters suggest that a table-top device should be able to detect
a photon rest mass at the level of $9\times10^{-50}$ grams, and measure
deviation in the exponent of Coulomb's inverse-square law at the level
of a few times $10^{-22}$, both representing an improvement of several
orders of magnitude over current laboratory measurements. In addition,
the apparatus would be immune to effects related to the modification
of the field by the instrument used to measure it.

We acknowledge Ross Spencer for his assistance with numerical calculations.
This work was funded by BYU's Office of Research and Creative Activities.


\begin{thebibliography}{19}
\expandafter\ifx\csname natexlab\endcsname\relax\def\natexlab#1{#1}\fi
\expandafter\ifx\csname bibnamefont\endcsname\relax
  \def\bibnamefont#1{#1}\fi
\expandafter\ifx\csname bibfnamefont\endcsname\relax
  \def\bibfnamefont#1{#1}\fi
\expandafter\ifx\csname citenamefont\endcsname\relax
  \def\citenamefont#1{#1}\fi
\expandafter\ifx\csname url\endcsname\relax
  \def\url#1{\texttt{#1}}\fi
\expandafter\ifx\csname urlprefix\endcsname\relax\def\urlprefix{URL }\fi
\providecommand{\bibinfo}[2]{#2}
\providecommand{\eprint}[2][]{\url{#2}}

\bibitem[{\citenamefont{Goldhaber and Nieto}(1971)}]{Goldhaber71}
\bibinfo{author}{\bibfnamefont{A.~S.} \bibnamefont{Goldhaber}}
  \bibnamefont{and} \bibinfo{author}{\bibfnamefont{M.~M.} \bibnamefont{Nieto}},
  \bibinfo{journal}{Rev. Mod. Phys.} \textbf{\bibinfo{volume}{43}},
  \bibinfo{pages}{277} (\bibinfo{year}{1971}).

\bibitem[{\citenamefont{Tu et~al.}(2005)\citenamefont{Tu, Luo, and
  Gillies}}]{Tu2005}
\bibinfo{author}{\bibfnamefont{L.-C.} \bibnamefont{Tu}},
  \bibinfo{author}{\bibfnamefont{J.}~\bibnamefont{Luo}}, \bibnamefont{and}
  \bibinfo{author}{\bibfnamefont{G.~T.} \bibnamefont{Gillies}},
  \bibinfo{journal}{Rep. Prog. Phys.} \textbf{\bibinfo{volume}{68}},
  \bibinfo{pages}{77} (\bibinfo{year}{2005}).

\bibitem[{\citenamefont{Jackson}(1975)}]{Jackson1975}
\bibinfo{author}{\bibfnamefont{J.~D.} \bibnamefont{Jackson}},
  \emph{\bibinfo{title}{Classical Electrodynamics}}
  (\bibinfo{publisher}{Wiley}, \bibinfo{address}{New York},
  \bibinfo{year}{1975}), pp. \bibinfo{pages}{5--9, 597--601},
  \bibinfo{edition}{2nd} ed.

\bibitem[{\citenamefont{Greiner and Reinhardt}(1996)}]{Greiner1996}
\bibinfo{author}{\bibfnamefont{W.}~\bibnamefont{Greiner}} \bibnamefont{and}
  \bibinfo{author}{\bibfnamefont{J.}~\bibnamefont{Reinhardt}},
  \emph{\bibinfo{title}{Field Quantization}} (\bibinfo{publisher}{Springer},
  \bibinfo{year}{1996}).

\bibitem[{\citenamefont{Arkani-Hamed et~al.}(1998)\citenamefont{Arkani-Hamed,
  Dimopoulos, and Dvali}}]{Arkani1998}
\bibinfo{author}{\bibfnamefont{N.}~\bibnamefont{Arkani-Hamed}},
  \bibinfo{author}{\bibfnamefont{S.}~\bibnamefont{Dimopoulos}},
  \bibnamefont{and} \bibinfo{author}{\bibfnamefont{G.}~\bibnamefont{Dvali}},
  \bibinfo{journal}{Phys. Lett. B} \textbf{\bibinfo{volume}{429}},
  \bibinfo{pages}{263} (\bibinfo{year}{1998}).

\bibitem[{\citenamefont{Kostelecky and Samuel}(1991)}]{Kostelecky1991}
\bibinfo{author}{\bibfnamefont{V.~A.} \bibnamefont{Kostelecky}}
  \bibnamefont{and} \bibinfo{author}{\bibfnamefont{S.}~\bibnamefont{Samuel}},
  \bibinfo{journal}{Phys. Rev. Lett.} \textbf{\bibinfo{volume}{66}},
  \bibinfo{pages}{1811} (\bibinfo{year}{1991}).

\bibitem[{\citenamefont{Elliott}(1966)}]{Robinson1769}
\bibinfo{author}{\bibfnamefont{R.~S.} \bibnamefont{Elliott}},
  \emph{\bibinfo{title}{Electomagnetics}} (\bibinfo{publisher}{McGraw-Hill},
  \bibinfo{address}{New York}, \bibinfo{year}{1966}), pp.
  \bibinfo{pages}{100--101}.

\bibitem[{\citenamefont{Cavendish}(1879)}]{Cavendish1773}
\bibinfo{author}{\bibfnamefont{H.}~\bibnamefont{Cavendish}},
  \emph{\bibinfo{title}{The Electrical Researches of the Honourable Henry
  Cavendish}} (\bibinfo{publisher}{Cambridge University Press},
  \bibinfo{address}{Cambridge}, \bibinfo{year}{1879}), pp.
  \bibinfo{pages}{104--13}.

\bibitem[{\citenamefont{Maxwell}(1873)}]{Maxwell}
\bibinfo{author}{\bibfnamefont{J.~C.} \bibnamefont{Maxwell}},
  \emph{\bibinfo{title}{A Treatise on Electricity and Magnetism 3rd edn}}
  (\bibinfo{publisher}{Dover}, \bibinfo{year}{1873}).

\bibitem[{\citenamefont{Williams et~al.}(1971)\citenamefont{Williams, Faller,
  and Hill}}]{Williams1971}
\bibinfo{author}{\bibfnamefont{E.~R.} \bibnamefont{Williams}},
  \bibinfo{author}{\bibfnamefont{J.~E.} \bibnamefont{Faller}},
  \bibnamefont{and} \bibinfo{author}{\bibfnamefont{H.~A.} \bibnamefont{Hill}},
  \bibinfo{journal}{Phys. Rev. Lett.} \textbf{\bibinfo{volume}{26}},
  \bibinfo{pages}{721} (\bibinfo{year}{1971}).

\bibitem[{\citenamefont{Crandall}(1983)}]{Crandall83}
\bibinfo{author}{\bibfnamefont{R.~E.} \bibnamefont{Crandall}},
  \bibinfo{journal}{Am. J. Phys.} \textbf{\bibinfo{volume}{51}},
  \bibinfo{pages}{698} (\bibinfo{year}{1983}).

\bibitem[{\citenamefont{Gronniger et~al.}(2005)\citenamefont{Gronniger,
  Barwick, Batelaan, Savas, Pritchard, and Cronin}}]{MetalizedGratings}
\bibinfo{author}{\bibfnamefont{G.}~\bibnamefont{Gronniger}},
  \bibinfo{author}{\bibfnamefont{B.}~\bibnamefont{Barwick}},
  \bibinfo{author}{\bibfnamefont{H.}~\bibnamefont{Batelaan}},
  \bibinfo{author}{\bibfnamefont{T.}~\bibnamefont{Savas}},
  \bibinfo{author}{\bibfnamefont{D.}~\bibnamefont{Pritchard}},
  \bibnamefont{and} \bibinfo{author}{\bibfnamefont{A.}~\bibnamefont{Cronin}},
  \bibinfo{journal}{Appl. Phys. Lett.} \textbf{\bibinfo{volume}{87}},
  \bibinfo{pages}{124104} (\bibinfo{year}{2005}).

\bibitem[{\citenamefont{Gustavson et~al.}(1997)\citenamefont{Gustavson, Bouyer,
  and Kasevich}}]{Gustavson97}
\bibinfo{author}{\bibfnamefont{T.~L.} \bibnamefont{Gustavson}},
  \bibinfo{author}{\bibfnamefont{P.}~\bibnamefont{Bouyer}}, \bibnamefont{and}
  \bibinfo{author}{\bibfnamefont{M.~A.} \bibnamefont{Kasevich}},
  \bibinfo{journal}{Phys. Rev. Lett.} \textbf{\bibinfo{volume}{78}},
  \bibinfo{pages}{2046} (\bibinfo{year}{1997}).

\bibitem[{\citenamefont{Christensen et~al.}(2006)\citenamefont{Christensen,
  Neyenhuis, and Durfee}}]{Christensen06nc}
\bibinfo{author}{\bibfnamefont{D.}~\bibnamefont{Christensen}},
  \bibinfo{author}{\bibfnamefont{B.}~\bibnamefont{Neyenhuis}},
  \bibnamefont{and} \bibinfo{author}{\bibfnamefont{D.~S.}
  \bibnamefont{Durfee}}, \bibinfo{journal}{arXiv:physics/0609128}
  (\bibinfo{year}{2006}).

\bibitem[{\citenamefont{Gustavson et~al.}(2000)\citenamefont{Gustavson,
  Landragin, and Kasevich}}]{Gustavson00}
\bibinfo{author}{\bibfnamefont{T.~L.} \bibnamefont{Gustavson}},
  \bibinfo{author}{\bibfnamefont{A.}~\bibnamefont{Landragin}},
  \bibnamefont{and} \bibinfo{author}{\bibfnamefont{M.}~\bibnamefont{Kasevich}},
  \bibinfo{journal}{Class. Quantum Grav.} \textbf{\bibinfo{volume}{17}},
  \bibinfo{pages}{2385} (\bibinfo{year}{2000}).

\bibitem[{\citenamefont{Lu et~al.}(1996)\citenamefont{Lu, Corwin, Renn,
  Anderson, Cornell, and Wieman}}]{LVIS}
\bibinfo{author}{\bibfnamefont{Z.~T.} \bibnamefont{Lu}},
  \bibinfo{author}{\bibfnamefont{K.~L.} \bibnamefont{Corwin}},
  \bibinfo{author}{\bibfnamefont{M.~J.} \bibnamefont{Renn}},
  \bibinfo{author}{\bibfnamefont{M.~H.} \bibnamefont{Anderson}},
  \bibinfo{author}{\bibfnamefont{E.~A.} \bibnamefont{Cornell}},
  \bibnamefont{and} \bibinfo{author}{\bibfnamefont{C.~E.}
  \bibnamefont{Wieman}}, \bibinfo{journal}{Phys. Rev. Lett.}
  \textbf{\bibinfo{volume}{77}}, \bibinfo{pages}{3331} (\bibinfo{year}{1996}).

\bibitem[{\citenamefont{Deslauriers et~al.}(2006)\citenamefont{Deslauriers,
  Olmschenk, Stick, Hensinger, Sterk, and Monroe}}]{Deslauriers06sa}
\bibinfo{author}{\bibfnamefont{L.}~\bibnamefont{Deslauriers}},
  \bibinfo{author}{\bibfnamefont{S.}~\bibnamefont{Olmschenk}},
  \bibinfo{author}{\bibfnamefont{D.}~\bibnamefont{Stick}},
  \bibinfo{author}{\bibfnamefont{W.~K.} \bibnamefont{Hensinger}},
  \bibinfo{author}{\bibfnamefont{J.}~\bibnamefont{Sterk}}, \bibnamefont{and}
  \bibinfo{author}{\bibfnamefont{C.}~\bibnamefont{Monroe}},
  \bibinfo{journal}{Phys. Rev. Lett.} \textbf{\bibinfo{volume}{97}},
  \bibinfo{pages}{103007} (\bibinfo{year}{2006}).

\bibitem[{\citenamefont{Luo et~al.}(2003)\citenamefont{Luo, Tu, Hu, and
  Luan}}]{Luo2003}
\bibinfo{author}{\bibfnamefont{J.}~\bibnamefont{Luo}},
  \bibinfo{author}{\bibfnamefont{L.-C.} \bibnamefont{Tu}},
  \bibinfo{author}{\bibfnamefont{Z.-K.} \bibnamefont{Hu}}, \bibnamefont{and}
  \bibinfo{author}{\bibfnamefont{E.-J.} \bibnamefont{Luan}},
  \bibinfo{journal}{Phys. Rev. Lett.} \textbf{\bibinfo{volume}{90}},
  \bibinfo{pages}{081801} (\bibinfo{year}{2003}).

\bibitem[{\citenamefont{Lakes}(1998)}]{Lakes1998}
\bibinfo{author}{\bibfnamefont{R.}~\bibnamefont{Lakes}},
  \bibinfo{journal}{Phys. Rev. Lett.} \textbf{\bibinfo{volume}{80}},
  \bibinfo{pages}{1826} (\bibinfo{year}{1998}).

\end{thebibliography}
\end{document}